# User Satisfaction-Driven Bandwidth Allocation for Image Transmission in a Crowded Environment

*Sandipan Choudhuri[1], Kaustav Basu[2] and Arunabha Sen[3] School of Computing, Informatics and Decision Systems Engineering at Arizona State University, Tempe, AZ 85281*
*{[1]s.choudhuri, [2]kaustav.basu, [3]asen}@asu.edu*

## 1. Introduction

We have moved on from the days when social media were dominated by texts; the popular social networking enterprises certainly have changed over the years, and their profound usage of digital images will continue to increase as time progresses. With the advent of new technologies, mobile phone image quality has made considerable progress. Consequently, the digital images produced by them occupy larger storage space. Due to these large sizes, images take larger bandwidth and more time for upload or download through the Internet, thus making it inconvenient for file sharing, especially in crowded environments. To combat this problem, in this paper we lay out a proposal for a user satisfaction driven bandwidth allocation scheme for image transmission in a crowded environment. The crowded environment that we have in mind includes among others, large indoor/outdoor concerts, soccer/football games and political rallies. A large number of individuals attend such events and a significant portion of them take images of the happenings in the arena with their cell phones and transmit them to their friends off the arena using apps such as Whatsapp or post them on Facebook. The bandwidth available in the events arena is usually constrained by the limited number of gateways to carry the data traffic from the arena to the external world. The users also expect that the transmitted images are delivered to their destination or posted in platforms, such as Facebook, in real time. Failure of the network service provider to ensure this requirement causes user dissatisfaction, which the network service provider clearly wants to avoid. We argue that *user satisfaction* is a function of the *received image quality* and the *delay* in receiving the image. The user satisfaction will be high if the received image is of high quality and arrived at the intended destination after a small delay. The image quality in turn depends on the *image size* ($x \times y$ inches) and *image resolution* ($z$ dpi) and a third parameter referred to as the *saliency concordance metric*, which is explained in detail in the section IV. The image *file size* is given by the product of the image size and the image resolution. Higher image quality can be obtained by having large image size and/or high image resolution. However, high image quality also implies large image file size. If the bandwidth allocated to a user by the network service provider is fixed, then the delay encountered by the user will also increase with increase in the image file size. Thus, the parameters, image quality and delay, essentially are antagonistic with respect to each other in their role in improving *user satisfaction*. User satisfaction is not only a function of image quality and delay but also a function of the economic aspect of the service provided by the operator (service charges). However, for the purpose of this study, we focus primarily on the technical component of the user satisfaction (image quality and delay) instead of the economic component.

The network service provider (NSP) would like *maximize* the number of satisfied customers. We assume that each customer $C_i$, $1 \leq i \leq n$ has a *customer satisfaction threshold* $\tau_i$, $1 \leq i \leq n$. However, due to limited *radio resources* (frequency spectrum, bandwidth) available to the NSP, it may not be possible for the NSP to provide services at a level that will satisfy all the customers. In such a scenario, the NSP has to carefully develop a *policy* to serve the customers so that *overall user dissatisfaction* is *minimized* or *overall user satisfaction* is *maximized*.

User-satisfaction-based bandwidth allocation schemes in wireless environment have been studied in [1, 10, 11, 12, 14]. Among these studies, the one that comes closest to the problem being discussed in this paper is [10]. However, there exists significant differences between the approach taken in [10] and the one proposed in this paper. In many studies on user-satisfaction-based resource allocation problem (including [10]), user-satisfaction is modeled over logarithmic or exponential functions. We propose a machine-learning-based approach to compute the user-satisfaction function. We plan to develop this function based on customer survey data that includes the *quality of image*, *delay encountered* and corresponding *user-satisfaction index*. The user-satisfaction function used in [10] does not take into account any *semantic information from the image being transmitted*. A major difference between our approach and the one proposed in [10] is that we incorporate *saliency-based information* in our image compression framework.

The authors in [1] study *satisfaction maximization problem* for both *real time* and *non-real time* services. They use two different approaches, one *heuristic* and the other *utility based* to find solutions to the satisfaction maximization problem in real time and non-real time environments. As correctly noted in [1], one of the most important aspect of NRT service is *information integrity*, i.e., information loss is *unacceptable*. However, the application scenario that's under consideration in this paper, i.e., image transmission in a crowded environment, a certain amount of information loss (either through image compression or any other means) may be *acceptable*, as long as it doesn't significantly degrade the image quality. A number of techniques are currently available for image compression [2, 5]. A short review of these techniques is presented in section III. In this paper, we propose a novel saliency-based image compression technique to reduce the image file size without significantly reducing user satisfaction.

The images transmitted from the crowded environments under consideration in this paper will have some *salient objects*. For example, in an image of a soccer match, it is highly likely that the viewer will be more interested in the player and soccer ball movements, rather than other objects in the image. This implies that some objects in the image will be considered more important (salient) than others. This importance will be based on a number of visual, as well as, cognitive factors. The recipients of the images will be more interested in these objects than other parts of the image. Accordingly, as long as the quality of the salient objects in the image remains high, it's expected that the user satisfaction will also remain high, irrespective of the quality of the non-salient objects.

Our proposed scheme of user satisfaction driven bandwidth allocation for image transmission takes advantage of salient object detection effort in the image processing domain [7, 8, 5, 3, 6]. We recognize that at the present time, due to privacy issues, the NSPs are unable to see the image and as such cannot run any salient object detection algorithm on it. However, we envisage a scenario where the users may be willing to provide a waiver to the NSP to carry out *salient object detection algorithm* on the transmitted image, particularly if it contributes to enhanced user satisfaction. In the following section we explain our scheme in detail.

## 2. User Satisfaction Driven Bandwidth Allocation Scheme for Image Transmission

In our model, we assume that n customers, $C_i$, $1 \leq i \leq n$, are attempting to transmit one image each, $I_i, 1 \leq i \leq n$, simultaneously. The size of image $I_i$, $1 \leq i \leq n$, ($x \times y$ inches), is denoted by $S_i$ and the resolution of $I_i$ ($z$ dpi) is denoted by $R_i$. The size of the corresponding image file, denoted $F_i$ is given as the product of $S_i$ and $R_i$, i.e., $F_i = S_i \times R_i$. The NSP controls one or more *gateways* near the event venue, whose combined bandwidth is denoted by B. We assume that each customer $C_i$ has a declared *Acceptable User Satisfaction Threshold* $AUST_i$ ($\tau_i$). We also assume that the NSP has some customer profile information from which $\tau_i$ for customer $C_i$ can be estimated. The NSP needs to adopt a policy to decide on the amount of bandwidth to allocate to customer $C_i$ (denoted by $A_i$) so that satisfaction of customer $C_i$ should have *minimal* deviation from the threshold $\tau_i$. Suppose that the bandwidth required by the customer $C_i$ to attain satisfaction threshold $\tau_i$ is $SA_i$. Due to fixed (and limited) bandwidth B that is available to the NSP, it is conceivable that at times $\sum_{i=1}^{n} SA_i$ will be greater than B. In such circumstances, the NSP has to allocate bandwidth $A_i$ to customer $C_i$, where $A_i \leq SA_i$. Clearly, such an allocation will cause *user dissatisfaction*. The goal of the bandwidth allocation policy of the NSP is to *minimize* user dissatisfaction.

In order to realize the NSP's objective of minimizing user dissatisfaction (or maximizing user satisfaction), one first needs a quantifiable way of measuring user satisfaction. In our model user satisfaction is a function of *quality* of the received image and the *delay* encountered while receiving the image. We denote the quality of image $I_i$ as $IQ_i$ and *delay* encountered by $I_i$ as $\delta_i$. The user satisfaction of customer $C_i$ is denoted by $US_i$ is taken to be a function of $IQ_i$ and $\delta_i$, i.e.,

$$US_i = f_1(IQ_i, \delta_i) \qquad (1)$$

The quality of image $IQ_i$ is a function of image size $S_i$, resolution $R_i$ and a third parameter referred to as the *saliency concordance metric* ($SCM_i$) (see section IV).

$$IQ_i = f_2(S_i, R_i, SCM_i) \qquad (2)$$

The delay encountered by customer $C_i$ is a function of the image file size $F_i$ and the bandwidth allocated to $C_i$ by the NSP ($A_i$) and is obtained by dividing $F_i$ by $A_i$:

$$\delta_i = f_3(F_i, A_i) = \frac{F_i}{A_i} \qquad (3)$$

## 3. Review of Image Compression Techniques

In this section, we first review saliency-based and non-saliency-based standard image compression techniques. This is followed by our proposed saliency-based user satisfaction scheme.

*3.1. Non-Saliency-Based Standard Image Compression*

Usually image compression is generally guided by standard compression schemes like JPEG (Joint Photographic Experts Group) and JPEG2000 which are independent of image context information. The JPEG algorithm operates on the YCbCr color space, where Y channel contains luminance information and Cb and Cr channels carry color information. As our visual system is more sensitive to luminance than color, the color information from the two channels are down-sampled in both the horizontal and vertical directions. This is followed by partitioning of each image channel into blocks of *8×8* pixels. Compression is applied on each image block on all the three channels. The technique initiates by applying Discrete Cosine Transformation over a *8×8* block, yielding a *8×8* matrix of DCT coefficients as a replacement of the original 64 pixel values. Based on the image quality demanded, the compression algorithm selects an appropriate quantization table of size *8×8*. This table is used to divide the DCT coefficient matrix values. In order to favor small gradual tonal changes over high-frequency transitions, the division operation rounds off to the nearest integer. A consequence of this quantization step is reduction of the variation in matrix values- the resulting matrix values will likely have same values or be zeros. Representation of such matrices having numbers that are either similar or zero is significantly effective for compression. The process terminates by compressing these coefficients by an arithmetic/Huffman encoding scheme, which leads to further reduction of file size.

The JPEG2000 standard [2] compresses images in a similar way as the JPEG format, with a major change in the transformation step- the usage of Discrete Wavelet Transform (DWT) in place of DCT yields better image quality at very high compression efficiency.

*3.2. Saliency-Based Image Compression*

Recognition of contextual cues aids in capturing underlying inter-relationships between objects of particular interest from a given image, and forms a task of paramount significance in the image processing and computer vision community. Consequently, automated identification and localization of these visually interesting regions, consistent with human perception, finds a wide range of applications. Along with extracting semantically meaningful descriptions of objects, exclusive processing of salient objects yields a compact region-based description of an image. Standard compression schemes like JPEG, JPEG2000 do not capture these semantic information, and severely degrade salient region data. This can be circumvented by utilizing image compression techniques that prevent the degradation of salient regions at low bit-rates [3, 4], and also reduce the cost of image storage and transmission. The filtered regions of an image can be explicitly handled, transmitted and re-constructed at the receiver end.

Being a classical problem of computer vision, salient object detection have been extensively studied in the last few years [7, 6, 5]. In [5], the authors proposed a new compression technique and used context-aware object detection framework proposed in [6, 7]. In our scheme, we use a novel salient object detection technique, which is distinctly different from the one proposed in [6, 7]. This scheme is discussed in detail in the next section. For the sake of completeness, we provide a short review of the saliency-guided compression scheme used in [5]. The technique progresses along two paths:

*Generation of wavelet saliency map:* The salient-object identification initiates with context-aware object detection [6, 7], which locates regions with different degrees of saliency. After object detection in a given image (through implementation of context-aware algorithm similar to Cadena et. al. [8]), we use a new technique to identify the salient regions with different levels of importance, which described in the next section. The magnitude of importance of a particular region is portrayed by its pixel values in the saliency map, with higher pixel values signifying higher saliency. In order to avoid the overhead of processing arbitrarily-shaped regions of interest in the saliency map, each region is approximated with a rectangular bounding box having a constant saliency value. In the subsequent stage, a transformation from the rectangular spatial saliency map to a wavelet domain saliency map is performed, so as to choose the coefficients that will be transmitted first.

For a wavelet coefficient, the wavelet saliency is defined as the summation of pixel values at all those points which have non-zero values of wavelet basis function. For a given image I and the corresponding spatial saliency map $s_I$, the original image dimensions is resized to M and N, where M, $N \equiv 0 \pmod{2^K}$ and K being the number of levels in the desired wavelet decomposition. The following equation is used to recursively compute the wavelet saliency $s_w^k$ for Low-Low (LL) bands using Haar wavelet transform, where k denotes the decomposition level $(k \in \{0,1,2,...,K-1\})$:

$$s_w^{k+1}(i,j) = \sum_{i'=2i-1}^{2i} \sum_{j'=2j-1}^{2j} s_w^k(i',j') \qquad (4)$$

Here $i = 1, 2, \ldots \frac{M}{2^k}$ and $j = 1, 2, \ldots \frac{N}{2^k}$. The basis of the recursion in equation (1) is $s_w^0(i,j) = s_I(i,j)$, where $s_I(i,j)$ is the spatial saliency at (i,j). In order to expedite the wavelet transformation process, the wavelet saliency values of Low-Low (LL) band is copied to the corresponding locations of the Low-High (LH), High-Low (HL) and High-High (HH) bands. Now, with a spacial saliency value of 1 for the background, the wavelet saliency values will be in multiples of $4k$ for $k \in \{1,2,...,K\}$ (since Haar wavelet's support is 4 pixels). Now, if a spatial saliency value of $4k'+1$ is used $(k' \in N)$ for salient regions, $k'$ levels of salient region wavelet coefficients will be transmitted before the $K^{th}$ level of background wavelet coefficients. In other words, the salient regions are k levels ahead from the background in terms of saliency.

*Generation of quantized wavelet transform coefficients:* This module initiates by performing Haar wavelet transform on the YCbCr image equivalent of the original RGB image. As the human visual system is more sensitive to luminance than chrominance, sub-sampling of both the Cb and Cr channels are performed, both horizontally and vertically, by assigning the LH, HL and HH wavelet coefficients to zero at the best scale.

Quantization of these wavelet transform coefficients are obtained on a sub-band basis. After being mean-subtracted, each sub-band is scaled to an 8-bit integer. A part of the coefficients is then chosen for transmission using the first step, as indicated by bitrate requirements. The selected coefficients, sequentially ordered by saliency, are written into a binary file. This file is then subjected to two-fold entropy coding- the first technique being LZ77 algorithm, followed by Markov chain based range coding [9]. The resulting file is transmitted as per the bit-rate constraints, along with the mean and scaling factors of the selected coefficients.

The received bounding box data is utilized to reproduce the original saliency map. An identical wavelet transform algorithm is deployed to compute the equivalent wavelet saliency map. Using these information, the decoder decodes the wavelet transform coefficients and assigns them in accurate locations in the wavelet decomposition structure. Re-scaling of each coefficient for every sub-band is performed, using corresponding scale coefficient and mean value, followed by an inverse wavelet transform. The resultant image will be in YCbCr color space, which is further subjected to transformation in the RGB color space to reconstruct the sent image. Figure 1 shows a reconstructed image for JPEG, JPEG-2000 and the saliency-based compression algorithm presented in [5] (image obtained from the UMD faces dataset).

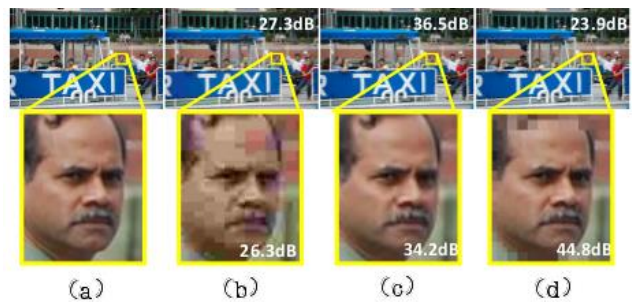

**Figure 1:** (a) Original Image, (b) Reconstructed JPEG image, (c) Reconstructed JPEG-2000 image, (d) Reconstructed image using the compression algorithm in [5] (Picture courtesy [5]).

## 4. Saliency-Based User Satisfaction Scheme

As noted earlier, user satisfaction is a function of image quality and delay. Image quality in turn is a function of image size, image resolution and *saliency concordance metric*. High user satisfaction is ensured by high image quality and low delay. Both image quality and delay are dependent in image file size. In general, high quality image increases the image file size, which in turn increases transmission delay. Since user satisfaction depends on both image quality and delay, the NSP has to carefully balance conflicting requirements of image quality and delay to maximize user satisfaction. The underlying assumption of the saliency based user satisfaction scheme is that as long as the quality of the salient objects in the image is high, user satisfaction will also remain high, irrespective of the quality of the non-salient objects. Various techniques for extraction of *salient objects* from an image are known [5, 7, 6]. The image Ii may comprise of object set $O_i$ and is given as $O_i=\{O_{i,1},O_{i,2},...,O_{i,p_i}\}$. In order to reduce the file size, the NSP sends only $q_i$, ($q_i < p_i$) *most salient objects* in Oi to the receiver and the receiver reconstructs the image with this set of objects. In order for the NSP to send the qi most salient objects, the NSP first has to *rank* the objects in the set Oi in order of their *saliency* from the perspective of the customer. The NSP most likely will not accurately know the customer perspective of the objects from the saliency standpoint. However, the NSP can *estimate* the customer perspective with a certain degree of confidence if the NSP has some knowledge of customer preference. We assume the NSP maintains a repository of images transmitted by the user over a period of time to understand that user's preference and perspective. The image repository of customer Ci is denoted by $IR_i$ and comprises of images $IR_{i,1}$, $1 \leq i \leq r_i$, i.e., $IR_i = \{IRi,_1, IRi,_2, ..., IR_{i,r_i}\}$. Let $IR_i^k \subseteq IR_i$ denote the subset of images in the image repository $IR_i$ that contains the object $O_{i,k}$, $1 \leq k \leq p_i$, an object in the set Oi. Let $C(I_i, IR_i^k)$ be the image in the set $IR_i^k$ that is *closest* to the image $I_i$ according to some *proximity/similarity* measure, such as Spatial Pyramid Matching [13]. Let $CIS(I_i) = \cup_{k=1}^{p_i} C(I_i, IR_i^k)$ denote the set of images closest to $I_i$ that contains the object $O_{i,k}$, $1 \leq k \leq p_i$. The NSP's estimated *rank* of the object $O_{i,k}$ from the user perspective will be equal to the rank of the image $C(I_i, IR_i^k)$ in the set CIS($I_i$) in terms it's proximity to the image Ii according to some proximity measure [13].

Using the ranking function defined above, the salient objects in the set $O_i=\{Oi,_1,Oi,_2,...,O_{i,p_i}\}$ are ranked from the *highest* to the *lowest*. After this ranking, $q_i$ highest ranked objects are selected for transmission to the sender. The *Saliency Concordance Metric* of image $I_i$ is given as the ratio of $q_i$ to $p_i$, i.e., $SCM_i = \frac{q_i}{p_i}$.

Earlier we stated that the quality of image $I_i$ $(IQ_i)$ is given as function of image size, resolution and *Saliency-Concordance Metric*. Now, we define the function $f_2(*, *, *)$ as :

$$f_2(S_i, R_i, SCM_i) = \omega_1 \frac{S_i'}{S_i} + \omega_2 \frac{R_i'}{R_i} + \omega_3 SCM_i \tag{5}$$

where $\omega_1$, $\omega_2$ and $\omega_3$ represents *weights* associated with *normalized* size and *normalized* resolution and saliency concordance metric $SCM_i$, with $\omega_1+\omega_2+\omega_3=1$. $S_i'$ and $R_i'$ represent the size and the resolution of the original image supplied by the sender and $S_i'$ and $S_i'$ represent the size and the resolution of the image actually transmitted by the NSP, after transforming the image through the salient object detection and modification algorithm.

Once the size $S_i'$ and resolution $S_i'$R'i of the image $I_i$ is determined, it's file size $FS_i'$ is also determined. The delay $\delta_i$ encountered for transmission of this file depends on the bandwidth $A_i$ allocated by the NSP for transmission of this file. As user satisfaction is given by the first equation, we need to identify the function $f_1(*, *)$ that maps a combination of $IQ_i$ and $\delta_i$ to $US_i$. We propose a machine learning based technique to determine the function $f_1(*, *)$. As indicated earlier, in our model the NSP maintains a repository of images that were sent by the user. This repository, in addition to the images, also stores the image quality $IQ_i$ and delay $\delta_i$ for each of the images. Moreover, in our model we assume that the NSP through questionnaire survey of the customer acquires the information regarding user satisfaction $US_i$ for each of the images transmitted. The information regarding $IQ_i$, $\delta_i$ and the corresponding $US_i$ can be maintained in a survey table. By applying appropriate machine learning algorithm, the function $f_1(*, *)$ can be computed with a high degree of accuracy.

We assume that each customer $C_i$ has a declared *Acceptable User Satisfaction Threshold* $AUST_i$ ($\tau_i$). When an image $I_i$ is transmitted by the NSP, the actual value user satisfaction $US_i$ corresponding to this image, may or may not exceed the $AUST_i$. If $US_i$ is smaller than $AUST_i$, the customer will be dissatisfied. In our model instead of treating user satisfaction as binary valued (satisfied or dissatisfied), we measure the *degree of user dissatisfaction* as the *difference* between $US_i$ and $AUST_i$. We conceive of two different objectives for the NSP. One objective may be to *minimize total dissatisfaction*, i.e.,

$$minimize \sum_{i=1}^{n} |US_i - \tau_i| \tag{5}$$

The second objective may be to *minimize the maximum dissatisfaction*, i.e.,

$$minimize \max_{i=1}^{n} |US_i - \tau_i| \tag{5}$$

The NSP has to minimize either one of the above two objectives, subject to the constraint that total allocated bandwidth to n customers cannot exceed the available bandwidth $B$, ie.,

$$\sum_{i=1}^{n} A_i \leq B \tag{5}$$

We propose the use of techniques such as Genetic Algorithms, Simulated Annealing or Tabu Search for finding the optimal bandwidth allocation that realizes the objectives, subject to available bandwidth constraint.

**5. Conclusion**

In this paper, we have presented a framework for saliency-based user satisfaction scheme. Our scheme relies on two key-features. First, to circumvent the biased realization of user-satisfiability with a generalized non-linear function, we propose a framework for quantifying user-satisfiability, based on customer profile. Second, in contrast to classical image compression technique, we propose a novel context-aware image compression scheme to reduce image file size. We are currently in the process of incorporating the ideas proposed in this paper in an experimental setup and evaluate the efficacy of our approach.

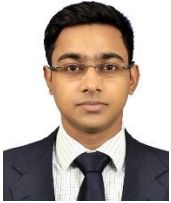
**Sandipan Choudhuri** received his Bachelor's and Master's degree in Computer Science and Engineering from West Bengal University of Technology and Jadavpur University, India, in the year 2013 and 2015 respectively. He worked as a Junior Research Fellow in the Computer Science and Engineering department of Jadavpur University from 2015 to 2017. He is currently a PhD student in Computer Science at Arizona State University. His current research interests include image processing and computer vision.

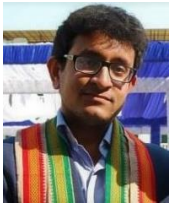
**Kaustav Basu** completed his Bachelor's and Master's degrees in Computer Science in 2014 and 2016 respectively. He is currently a PhD student in Computer Science at Arizona State University. His interests are the application of graph theory and machine learning, in the field of sports.

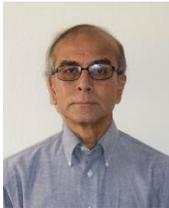
**Dr. Arun Sen** is a Professor of Computer Science and Engineering at Arizona State University. His expertise is in network modeling and optimization. He has over a hundred research publications in peer reviewed journals and conferences on various aspects of networks. In recent times his research is supported by the AFOSR, DTRA, ONR, ARO and the NSF. He served as an Associate Editor of the IEEE Transactions on Mobile Computing. He was a Visiting Scholar at University of California, Berkeley, University of California at San Diego, University of Passau in Germany and Wroclaw University of Science and Technology in Poland. In 2015, he was a Fulbright Fellow at INRIA at their North European Research Center in Lille, France. He started the International Workshop on Network Science for Communication Networks (NetSciCom) in 2009. The ninth edition of the workshop took place in conjunction with IEEE Infocom in 2017 in Atlanta.